# Person-centered and qualitative approaches to network analysis in physics education research

Adrienne L. Traxler[1], Camila Mani Dias do Amaral[2], Charles Henderson[3], Evan LaForge, Chase Hatcher[2], Madison Swirtz[2], Ramón Barthelemy[2]

[1]Department of Science Education, University of Copenhagen, Copenhagen, Denmark
[2]Department of Physics and Astronomy, University of Utah, Salt Lake City, UT, USA
[3]Department of Physics and Mallinson Institute for Science Education, Western Michigan University, Kalamazoo, MI, USA

## Abstract

Network analysis has become a well-recognized methodology in physics education research (PER), with study topics including student performance and persistence, faculty change, and the structure of conceptual networks. The social network analysis side of this work has focused on quantitative analysis of whole-network cases, such as the structure of networks in single classrooms. Egocentric or personal network approaches are largely unexplored, and qualitative methods are underdeveloped. In this paper, we outline theoretical and practical differences between two major network paradigms—whole-network and egocentric—and introduce theoretical frameworks and methodological considerations for egocentric studies. We also describe qualitative and mixed-methods approaches that are currently missing from the PER literature. We identify areas where these additional network methods may be of particular interest to physics education researchers, and end by discussing example cases and implications for new PER studies.

## I. Introduction

Network analysis has become recognized in physics education research (PER) as a lens for studying complex systems in terms of both entities and the relationships between them [1–3]. One way to categorize network analysis is by whether the network is social—with the nodes representing people—or whether the nodes represent other things like ideas. In the first case, a network framing offers tools for considering many-body situations such as classrooms, where individual outcomes are expected to arise in collaboration with others [3,4]. In the second case, networks may be used to study the structure and configurations between concepts, such as items in conceptual inventories [5] or the knowledge of physics teachers [6]. This paper focuses on the first case, or studies using social network analysis (SNA).

Another way to classify analyses is by whether the study concerns whole or personal networks. In whole-network approaches, a population boundary is defined (e.g. all students in a class section, all items on a concept inventory) and researchers seek a complete accounting of ties between this set of nodes. Egocentric or personal network analysis is a complementary framing that studies single actors and their surrounding connections. To date, almost all PER network studies have used whole-network approaches, and the primary aim of this paper is to bring egocentric approaches into the methodological domain of PER. Because many of these studies use qualitative methods, we also give significant attention to qualitative and mixed-methods network methodology.

In egocentric social network studies, the focus is on an individual actor—the ego—and mapping all of their connected people–-alters–-for one or more types of relationship. In many cases, the research also considers ties among alters, and this information comes from the ego rather than from the alters themselves. Multiple ego networks are typically collected and may be analyzed qualitatively [7,8], quantitatively [9,10], or with mixed methods [11,12]. Networks may overlap if the egos have alters in common, but this is not required, and in many cases the egos do not know each other at all. Unlike in whole-network studies, the goal is not to comprehensively sample one bounded space, but to learn about the size, structure, content, or functions of the personal networks of some type of person.

This approach offers new applications of networks in PER, but it uses different theoretical framings and sometimes different methods compared to whole-network studies. In this paper, we begin by briefly reviewing recent network analysis trends in PER and then introduce the methodological space that is not mapped by that work. We will lay out several key dimensions in the design of ego network studies, give a range of examples and resources, and highlight areas where the egocentric approach might broaden or deepen the use of network methods in PER. We highlight several educational studies with examples of egocentric and/or qualitative tools and end by discussing possibilities and implications for network studies in PER. One of the strengths and challenges of network analysis is that it spreads over many disciplinary and network-specific journals, so throughout the paper we have tried to include a diversity of resources that may not be familiar to Phys. Rev. PER readers.

## I.A. Terminology

Node: A person or other entity (e.g., a department or organization); networks are made of nodes connected by ties.

Tie: A common name in SNA for the connections between nodes in a network; also called links, connections, or edges (especially in quantitative studies).

Ego: The node at the focus of a personal network. This would be the person interviewed, surveyed, observed, etc. during data collection.

Alter: Other nodes in the network, connected to the ego.

Sociogram: Also known as "network diagram," a visual representation of the network showing the nodes and ties between them.

## I.B. Network analysis in PER: Recent trends

Network analysis has roots in both sociology and mathematics and lends itself to a rich diversity of research questions and contexts. In education, it has often been taken up because of its potential to describe complex systems of people and relationships at many scales [13,14]. However, reviews of networks in education have noted a dominance of quantitative descriptive studies which explore a relatively limited range of learning contexts or research questions [13,15,16]. In PER, the most common set of methods and study contexts share this potential for expansion.

To overview recent use of SNA in PER, we performed keyword searches [17] covering April 2017-November 2023 on ERIC, the Web of Science, Google Scholar, the proceedings from the Physics Education Research Conference (PERC), and the International Research Group on Physics Teaching (GIREP). We classified the relevant results by whether they used a whole or ego network approach and whether they used qualitative, quantitative, or mixed methods. Other possible categories (e.g., study population) were not distinguished because our primary interest lay in probing the methodological space.

Of the 21 papers we found, 18 used quantitative methods, 2 used mixed methods[1], and 1 used qualitative methods. All but one used a whole-network approach. To our knowledge, only two egocentric-focused PER network papers have been published: one by Goertzen and collaborators [20], before the period of this sample, and one by Wu and collaborators [18]. The themes explored by the sampled papers include function and structure of social networks [18,19,21–29], persistence in physics [25,30–32], academic performance [25,32–34], changes and development of social networks over time [18,28,32,35], assessing the effectiveness of specific programs [18,29,34,36], and theory [2,3,37]. One area that is largely absent from PER, but rapidly growing in other educational domains, is the combination of network analysis and learning analytics to study online or in-person learning environments [15,16].

Although the incorporation of SNA in PER is relatively recent [2], the method contributes to physics education by providing relevant tools and a theoretical perspective on how connections between students and their peers or instructors affect different aspects of education. SNA also has a wealth of methods that have not yet been applied to PER. In particular, qualitative and egocentric approaches are virtually unexplored despite their unique potential contributions to the field.

---

[1] Two papers that coded video data to generate networks [18,19] are grouped as quantitative rather than mixed-methods because the purpose of the coding was to count the presence or absent of network ties, not to describe details of the interaction.

## I.C. Framing egocentric vs. whole-network approaches

*Complete networks*, or *whole networks,* are derived from a sociocentric (as opposed to egocentric) approach to network analysis [38,39]. This approach is helpful to assess the position of all nodes simultaneously within a network and is used when the research aims to understand how a specific group is structured. It rests on the assumption that the researcher has access to the entire group and knows its boundaries [39]. A physics class [27] or a university department [40] are examples of complete networks in PER.

The egocentric approach focuses instead on the network of one specific node (ego) and enables comparisons between the networks of different egos [38,39]. Ego networks can be extracted from whole networks [38,41], but it is common to collect data from egos directly. This can be helpful when it is not possible to access the entire population of interest. Ego networks can also be a useful approach for networks that do not overlap, but where researchers still want to better understand the connections of members of different groups [39].

Figure 1 shows an example of the difference between these two approaches to social networks. The first panel might represent a classroom network collected in a typical PER study. The second panel shows an alternate focus, where the networks of selected students are highlighted for a focused study (one route for mixed-methods approaches, see [42,43]). More detailed data collection centered on those egos could include many people outside the class roster boundary such as teaching assistants, friends, or family members. Figure 2 shows those egos might have important ties not visible in the classroom network but salient from the egocentric point of view. The first panel shows the original classmate ties plus new non-classmate alters. The second panel adds information: alters are positioned according to closeness, node shapes are used to indicate role (e.g., circles for classmates and triangles for students in other courses), and line types indicate different ties (e.g. in-class interaction as solid line, and participation in a study group as dashed line).

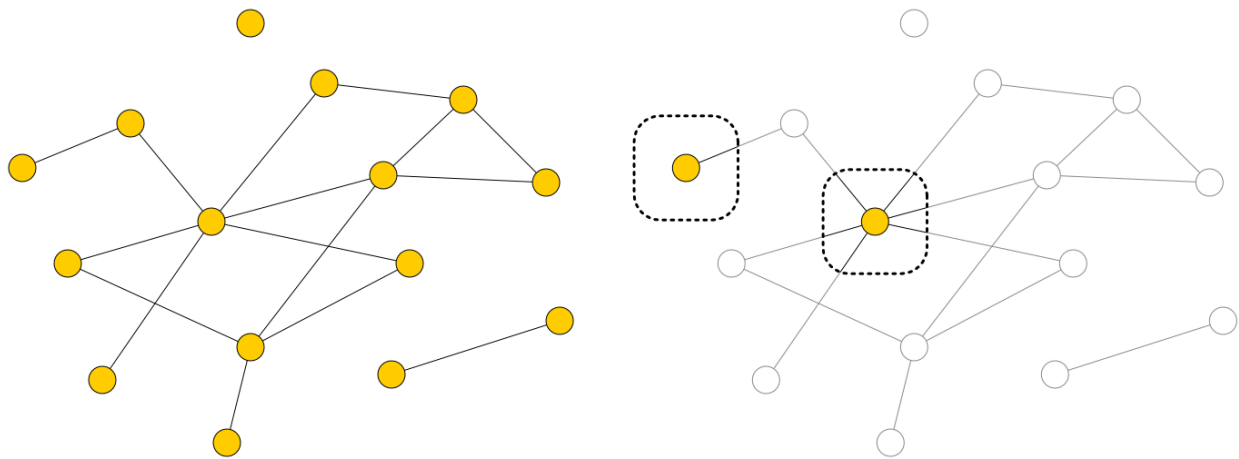

*Figure 1: A sample whole network (left) and possible ego network focus (right).*

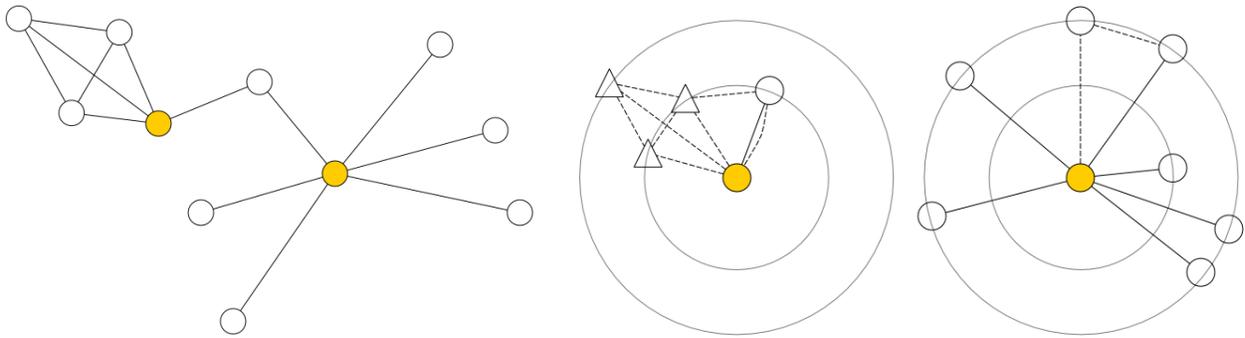

*Figure 2: Personal network connections for the two egos selected in the right panel of Fig. 1. On the left is the portion of the class network that connects to these two egos, plus additional alters from outside class (for the leftmost ego). On the right, the same set of alters is arranged by closeness and shapes are used to indicate alters' role. Dashed or solid lines denote separate types of ties, and both egos have ties that were not apparent from the original whole-class survey.*

Whole-network studies are often quantitative, reporting on a range of characteristics such as the distribution of links between actors, membership in sub-groups or communities, or the flow of resources or information through the network. These same characteristics may also be described qualitatively for smaller networks, for example to characterize the discussion structure of a class [44].

Egocentric studies are also used for quantitative research questions, for example about the variation in sizes or densities of personal networks, or the differences in tie strength between certain interaction types [9,18,19]. They offer additional range for qualitative or mixed-methods studies that focus on a "depth over breadth" approach. Ego networks are especially suited to investigations where the nuances of network relations are key, such as questions about the content or meaning (beyond binary existence) of ties. For example, a network study of physics majors in a department might ask egos who they do homework with, but also how they perceive the expertise of those people and whether the study group has been stable or changing over time. These details could be a valuable addition to work on identity development, where recognition is a key part of theory [45,46] and might be reflected in the topology of peer networks.

In both whole and egocentric network studies, it is axiomatic that the connections between people are fundamental to understanding their lived reality. Whole-network studies position themselves to some extent "above" (or at least outside) the group in question, in effect describing the forest instead of individual trees. Ego network studies are situated more in the perspective of an individual person, understanding the shared social reality as a collection of overlapping personal worlds. This person-centered approach can be taken up through a number of different theoretical frameworks and data collection tools, which we discuss next.

We organize the rest of the paper in terms of several "big questions" that might be asked when planning a research project. These questions are potentially useful in all network studies, but especially helpful in sketching a map of this newer-to-PER methodological space:
- What interests you about the network? (Sec. II)

- Why do you think the network is important? (Sec. III)
- How do you intend to study the network? (Sec. IV)

Section V discusses several examples from the educational literature in light of this material, and section VI concludes with recommendations for PER.

# II. Conceptualizing: What interests you about the network?

The first "big question" starts from curiosity: what interests you about the network? Studies can start from an overheard bit of conversation, an observation that some students in a class seem to gravitate towards each other, an experience of a close graduate school cohort, or many other threads. One brainstorming tool is to think about different possible dimensions of networks that can be foregrounded. Perry et al. [38 chapter 1] articulate four dimensions:
- *structure*, such as the network's size or patterns of ties;
- *function*, such as the type of exchanges or resources available in the network;
- *strength*, such as the relative durability or variety in bonds;
- and *content*, such as attitudes and beliefs present in the network.

All of these relate to the dual nature of network systems—the fact that they are built of both nodes and links—but focus on different facets of that duality. Many PER network analyses focus on structure, or functional questions in the case of more specific prompts such as homework help [26]. But the number of strong ties in students' departmental networks might be an important indicator for retention, or colleagues' views about teaching might be highly relevant for departmental reform efforts. Later decisions in the study design can be clarified by thinking around these dimensions and which one(s) to highlight.

# III. Theoretical frameworks: Why do you think the network is important?

Various theoretical frameworks have been used in social network studies, based on different mechanisms that might be important in the participants' context. In some cases, researchers are primarily interested in the effects of network structure on the actors, while in other cases, the focus is on what causes that structure to exist or change [47]. In terms of "big questions", this point can be framed as "Why do you think the network is important?" Below we expand on three possible answers to this question which have been well developed in the network literature [38].

## III.A. Social support: importance to well-being

The social support framework focuses on how network ties connect people to resources that sustain them in times of difficulty. These supports might be instrumental (practical assistance), emotional, appraisal (for example, of a problem), or monitoring [38 chapter 1]. In literature on educational or workplace networks, instrumental and emotional supports are the most commonly discussed. For example, Atherley and collaborators [48], in a mixed-methods study of transitions in medical student clerkships, found that both emotional and instrumental support ties changed greatly over the first four months. Peer ties grew especially prevalent due to the importance of venting and of sharing professional norms picked up during clerkships. In another example, Skahill [49] studied the social support networks of first-year students at a technical arts college and found that residential students experienced a sharp density drop and need to rebuild their networks compared to commuter students who had more local ties. However, residential students who formed new ties reported significantly greater feelings of success at school, and these new school-centered networks may have boosted their persistence.

Additional overview and examples of the social support literature for network studies is given by [38,50,51]. This framework might be especially useful for investigating retention and persistence (e.g., of students in a physics major) or difficult transitions (e.g. beginning a teaching career or graduate program). It can also help to clarify the kinds of resources that are important, for example in distinguishing multiple sources of support such as emotional versus instrumental or material connections.

## III.B. Social capital: importance to power and brokerage

Social capital, as used in network analysis, refers to access to resources in a social network. Lin [52] identifies three elements: the resources are part of a social structure (embeddedness), accessible to individuals (opportunity), and used in purposive actions (use). There is conceptual overlap between this framing and the social support literature discussed above. However, social support studies tend to focus on the well-being of network members, while social capital studies orient towards more external questions of power and brokerage. An influential work in the development of this area was Granovetter's "The strength of weak ties" [53], which discusses how important information such as new job leads may come from more tenuously-connected parts of personal networks, as opposed to the denser set of strong ties in a person's core relationships. One explanation for this trend is that peripheral network members are more likely to bring novel and non-redundant information. This idea has been extensively studied in workplace networks that explore concepts of brokerage and bridges, which connect people around "structural holes" [54].

In educational contexts, social capital has been used to frame students' resources and access at various ages. In one example, Rios-Aguilar and Deil-Amen [55] map and characterize social capital in the networks of Latina/o university students. By distinguishing nine categories of social ties such as family, clubs and organizations, and professional networks, they were able to identify how students moving from high school to university tended to have capital relevant to "getting in," but far fewer ties who could help with strategizing success in college or planning professional trajectories. Their analysis also revealed regions of networks where these ties could be sustained and strengthened, such as bridge programs.

Knaub et al. [56] suggest an alternate focus of social capital that can benefit education researchers, by focusing on the ties in academic departments as sites of instructional change.

They frame the central question of the social capital perspective as "what is being exchanged in a tie?" and note the importance of weak ties, access to new ideas, and brokerage in studying these social systems. In our own study of the professional support networks of women and LGBT+ physicists [57], social capital considerations helped to formulate which alters to elicit and what kind of ties to explore in the interviews.

## III.C. Social influence: importance to spreading information and attitudes

The social influence framework links network structure with actor attributes through processes such as peer pressure, persuasion, and competition [38,58]. Many PER network studies focus on classroom networks with the expectation that students will transmit information and attitudes through their connections, and social influence is one framework to explore these mechanisms. Some investigations place primary importance on the cohesiveness of local network ties, while others examine the equivalence of tie structure, positing that people in similar social circumstances will be driven to act in similar ways [58].

Researchers interested in building community in student groups may find useful tools in the network literature on homophily [59], which is the tendency for ties to form and persist among similar individuals. The mechanisms behind this tendency are nuanced by preference, availability, and influence [38 chapter 7], which can be studied qualitatively or quantitatively. Literature on structural holes [54] has probed the positive and negative constraints that arise from local network cohesion. This work has often focused on organizational and workplace settings, but also has been used to explain diffusion of ideas.

In one example, Lorenz et al. [60] review a number of sorting effects that tend to homogenize friendship networks, then use widespread sampling and a longitudinal design to distinguish peer selection effects from students' educational expectations. Beyond classroom content, physics education researchers might find this framework valuable to study topics like the cohesiveness of departmental networks, or how change initiatives propagate through students and faculty.

## III.D. Other frameworks

Our goal is not to argue for one theoretical framework over another, but to advocate for the importance of using some framework to guide study design. Clarity about study goals and theoretical constructs is key to designing valid data collection: as we discuss below, the wording of prompts to elicit connections can affect the size and membership of the network, even for questions that seem closely related [61].

For researchers, the key questions are: why do you think connections between people are important in this setting? What do you hope to understand better by mapping those connections? The theoretical framework of the study can highlight and focus these choices.

# IV. Methodology: How do you intend to study the network?

Most methods used for network data collection are used in other areas of PER as well. However, network-specific issues arise in collecting usable data and exploiting the relational perspective in analysis, as well as the unique tool of sociograms. We highlight key results here, beginning with a discussion of the different strengths of qualitative, quantitative, and mixed-methods designs.

## IV.A. Research approaches

### IV.A.1. Quantitative designs

SNA was rooted in quantitative sociology [2], and in PER this has often been supplemented by complex network studies taking an algorithmic viewpoint on analysis (e.g., [62]). For the purposes of this paper, a quantitative approach is characterized by the use of mathematical tools to map networks, measure their properties, or describe the position of nodes [63,64].

This orientation is common in whole-network studies and can be used in egocentric studies as well: for example to survey the size, homogeneity, level of resource access, or other network traits for a population of interest. The work of Lorenz and collaborators referenced above [60] is one example, and McCarty [9] argues for the use of ego network structural measures to distinguish socialization patterns. If an appropriate sampling strategy is used, statistical calculations can be used much more straightforwardly for ego networks than for whole networks, where the interdependence of the observations requires special methods [65].

One important decision is whether to include or remove the central node—the ego—when calculating structural measures. An example is Wu et al.'s work [19] where physics lab teaching assistants (TAs) were the focus of the network, with students represented at the group level. Retaining the ego (TA) allowed for comparing levels of TA-group interaction between lab curricula. If the focus was instead on levels of student-student talk in different implementations, it might make more sense to subtract instructor ties before analyzing the remaining links. This question doesn't really occur in whole-network studies, but is significant for egocentric designs because ego-to-alter ties will by definition be a substantial fraction of the network. In general, McCarty & Wutitch [66] argue for removing the ego for questions of network-on-ego effects, and keeping it for research that focuses on ego-on-network effects.

### IV.A.2. Qualitative designs

Qualitative designs can use many approaches toward inquiry (e.g. ethnography, narrative analysis, phenomenography), but are united by the importance they place on "the *understanding of meaning*" [8] and the nuances of lived reality. Though qualitative network methods are rare in PER, they are well developed in other fields. Hollstein [8] recommends qualitative analyses for probing the mechanisms or dynamics behind observed network structure, as well as for validating or understanding other results. Crossley and Edwards [11] expand on the theme of mechanisms, arguing that individual network results are cases that do not generalize, but that qualitative methods allow researchers to identify and explore broader mechanisms.

Qualitative methods are an essential precursor to many larger-scale quantitative studies and are an especially strong choice when the network dynamics are not well understood. But they are also the primary means to focus on the nuances of network members' social worlds: the subjective meaning of relationships and how they influence participants [7]. Because of this sensitivity to nuance, qualitative approaches are especially valuable for understanding multiplex ties or for exploring sensitive topics. In a physics context, this might mean mapping the ties of support and capital that students need over their degree program, or examining the inclusion and exclusion dynamics of study groups in the department. Understanding these processes or tendencies is key to tease apart the behavior of complex systems, such as classrooms or organizations, where simple causal claims are rarely possible.

To read more in this area, Hollstein [8] gives an overview of recommendations with many examples, primarily using grounded theory or ethnography. Another excellent introduction is work by Nimmon and Atherley [7 Fig. 1], directed at health professions education researchers and with many parallels in PER. Their "research onion" diagram outlines many study design decisions that are important to document and report such as theoretical frameworks, methodological choices, time dimension, and type(s) of data collected.

## IV.A.3. Mixed-methods designs

Mixed-methods social network analysis (MMSNA) has received growing attention in recent years. Froehlich et al. [12] define MMSNA as "any SNA study drawing from both qualitative and quantitative data or using qualitative and quantitative methods of analysis," where those methods are "thoughtfully integrated." Edwards [64] argues that mixed methods combine the strengths of quantitative SNA for systematic and precise mapping with the nuance of qualitative tools for understanding what that structure actually means. She discusses broad strategies that may be useful, such as using quantitative and qualitative approaches in sequence to focus the analysis. In one example, Martínez et al. [42] collected a large body of data from a computer-supported collaborative learning environment and used quantitative network analysis to highlight groups of students for more in-depth qualitative analysis. In another example, Atherley et al. [48] found statistically significant shifts in the types of support network alters named over time by their student participants, then returned to the qualitative interview data to interpret and explain these shifts.

Looking to future studies, Edwards [64] notes that mixed methods are especially powerful for understanding context and change. In PER, this could be at the level of an individual student or a department/program, and could take a whole-network or egocentric approach depending on the research questions. Because network objects inherently combine both structure and content, a blend of methods that addresses both at once can be very powerful, allowing a rich view of the mechanisms driving complex social systems [11,64].

To read more in this area, the overview by Edwards [64] is a good starting point. Many examples of specific methods combinations are given there as well as in Crossley and Edwards [11] and Froehlich et al. [12], which lays out critical questions about *why* methods are being mixed and what to report.

## IV.B. Data collection and analysis

### IV.B.1. Research approach influence

Quantitative data collection methods include distributing questionnaires about connections between nodes, observations, or using digital traces such as forum posting logs to indicate ties. Quantitative approaches to SNA are usually connected to a sociocentric approach, but it is also possible to select nodes within a whole network to focus on as subjects of qualitative research [8].

Qualitative approaches involve an overlapping set of data collection methods: observations, interviewing, and collecting documents or archival material are all common [8]. If actual behavior is the focus of the study, observation may be the most accurate tool (though see [67] for notes on difficulties). If perceived networks are the most relevant, which is often the case in social support studies, then interviews or surveys may give more salient data.

In the mixed-methods realm, Froehlich [43] mapped MMSNA studies in education and found that they were dominated by a small cluster of methods. Studies tended to use surveys or semi-structured interviews for data collection, followed by network visualization and/or calculating summary statistics. Froehlich's analysis also found that while data collection and analysis methods were generally reported in sufficient detail to classify, much less information was given about sampling procedures or data preparation. His visual map of method co-occurrence is a good resource to see what has been done, as well as highlighting potentially under-developed areas.

In PER, most network studies use surveys, where one or more prompts are used (often accompanied by rosters of names) to solicit network ties from respondents. The survey is given to all members inside the predefined network boundary, and the resulting whole network (typically with some missing data) is analyzed. In egocentric SNA, the process of soliciting alters is called "name generation," and follow-up questions to collect alter traits or ties comes from "name interpreter" questions. Both of these tasks often use surveys and interviews, and for interviews there is growing use of participant-drawn sociograms.

Because of their importance, we will focus on survey and interview data collection below, with extra attention to egocentric issues and specific notes about sociograms. For observational designs, Marsden [67,68] notes key studies and issues to consider.

## IV.B.2. Surveys and interviews for network data

Surveys have been a dominant data collection tool in SNA for many years [67–69]. Two considerations that are especially relevant for SNA studies are how memory can affect network representations and how the network is theorized and operationalized in the prompts.

Like all recall-based data collection, surveys are subject to memory biases, but also show reasonable accuracy and fidelity for many research purposes. Early studies often limited name generators to a small number of responses, such as five, to limit participant burden. However, this can seriously distort network results, so it is not recommended ( [68], though see [67] for sampling-based workarounds). Name generators with context, specific relationships, or recent time frames can aid recall. In general, people are better able to remember closer connections and core members of their networks [69,70]. If more tenuous connections are also important to the research question (often the case, see [53]), the core contacts may be helpful bases for follow-up prompts (e.g., "are there any other members of that study group?").

Name interpreters raise the question of whether reality or the respondent's perception of reality are most important. For example, when reporting more observable traits like demographic characteristics, ego's accuracy is generally good. But reports of alters' attitudes or beliefs tend to be slanted toward projecting the ego's own values [68], and this should be noted when making claims based on those responses.

The wording of the name generator can substantially affect the network. One common division of generators is by type: role, exchange, or interaction [68,71]. Role relationships such as "close friends" or "people you discuss important matters with" have often been used in sociology-based network research. Exchange generators target more specific dimensions, like "who are all the people you work with on physics homework?" or "who would you ask for advice about going to graduate school?" Unless only a very specific setting or only a very rough idea of the network are desired, multiple name generators are probably necessary to span the ties of interest [67]. Questions might also include negative support ties, which are seldom studied but can be very significant [68,71]. Finally, interaction-based name generators have also been explored in sociology, often in the context of attempting to gather the entirety of a person's contacts (perhaps over some threshold, like 5 minutes) in a time frame. Interaction prompts are often used in PER studies, for example giving students a roster and having them complete "I had a meaningful interaction with these people this week" [31].

To study support networks, van der Poel [72] advocates for exchange-oriented name generators. He argues that interaction networks often neglect content and importance of their contacts, role relations (e.g. friends) can vary in how each type of role matters to the person, and affective prompts (such as asking about closeness) can vary greatly in interpretation. The exchange approach does neglect currently-dormant connections, so it may underestimate the full support available. In a study of various prompts for emotional, instrumental, and social support, van der Poel [72] found that 3-5 name generators could span most of the variance and role relationships in the sampled personal support networks. That study, as well as comparative work by Milardo [71] and by Ferligoj and Hlebec [73] are good resources to learn more about name generator choices.

To return to the broader issue of framing and wording, one critical finding is that role, exchange, and interaction generators all produce different networks and should generally not be used as proxies for each other. In a comparative study [71], interaction and exchange networks had uncorrelated sizes and did not have substantial overlap of members. In a PER context, for example, having a robust in-class interaction network may not translate to having good access to help on the homework. Answers to the earlier "big questions" in study design can help to carefully choose targets for data collection.

### IV.B.3. Sociograms as data

Sociograms are a common tool in network interviews, and a small but growing number of studies also use them in computer surveys with no interviewer present [74]. Sociograms are a visual cue that often provokes extra recall or detail from participants. At a deeper level, they can become a thinking tool for the participant, taking them from listing to reflecting and theorizing about their relationships [75].

Sociograms vary in shape and level of structure or formality. A popular design is to give participants pre-drawn concentric circles with the ego at the center, and larger circles representing decreasing closeness [76]. Figure 3 shows an example design we used to collect data with Google Jamboards in a study of the professional support networks of women and LGBT+ physicists.

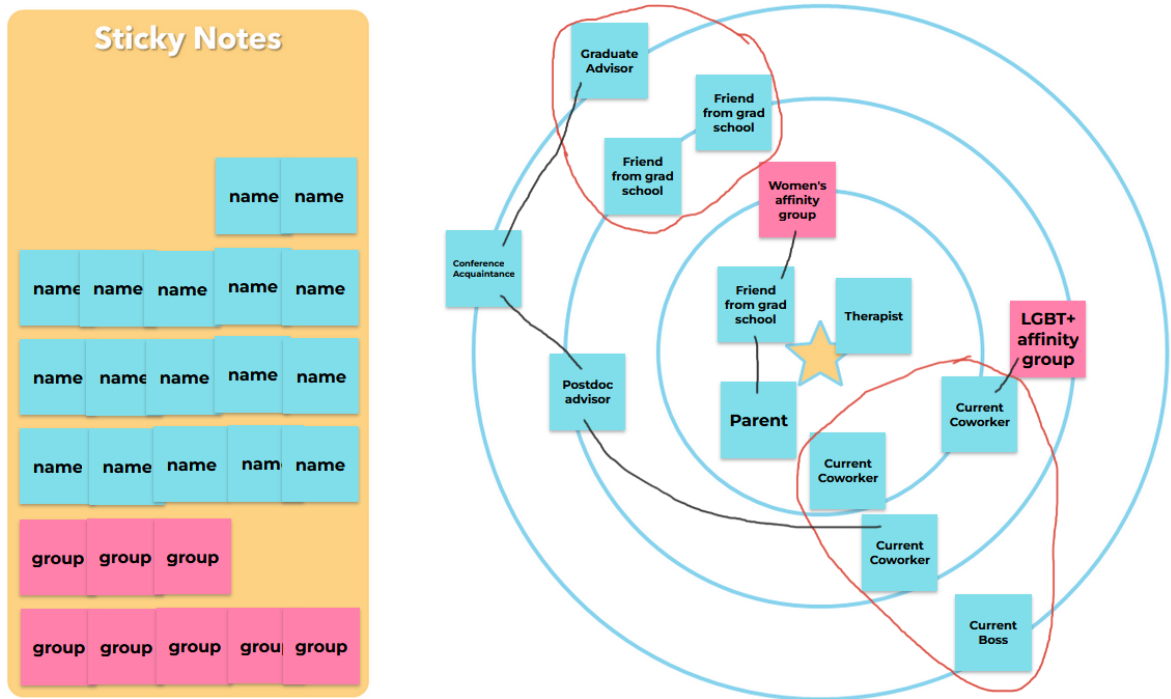

*Figure 3: Example sociogram a participant might construct during the interview. Blue sticky notes are for individuals and pink for groups. Names would typically be used by participants, but here are replaced by anonymous descriptors.*

Sectors may be pre-marked for roles such as friends or coworkers [74,75], or the plotting space may be free of such divisions, or may even be a completely blank free-drawing area [77]. The level of structure influences the kind of data that is collected and what it is possible to do with it. Hollstein and collaborators [77] compared several sociogram designs and found that the concentric circles format was the most popular with participants. The free-mapping variant produced the most quirky data, with network members including dead people and pets. The authors caution that if researchers plan to compare networks, a standardized script for introducing and using the diagram is important, because even a pre-printed structure such as circles is still subject to varying interpretations.

Ryan et al. [78] discuss other facets of visualizing networks in interviews. They found it to be an especially helpful tool to talk about dynamism and changing ties, which is often a struggle for survey-based quantitative network studies. However, they also found a few instances of participant discomfort, if people felt guilty about forgetting a contact or were uneasy at the status implied in ranking alters' distance from the center. Echoing some of Hollstein and collaborators' cautions, the researchers found that completion could be idiosyncratic, so detailed comparisons between sociograms should be made with care.

For practical considerations, Hogan et al. [79] introduce an interview method that combines survey-style name generation with placement on a sociogram. Their suggestions for marking alter-alter ties can greatly alleviate the response burden for participants, which otherwise increases dramatically as the number of alters grows [80]. Molina [81] has recommendations for question sequence to help recall of groups versus communities. Finally, if detailed knowledge of alter-alter tie structure is paramount, then a more traditional survey grid that requires explicitly evaluating every tie may be best, as visual mapping has some bias toward the most subjectively salient alters and ties [74]. As always, the nature of the tool affects the data collected, so researchers should consider in advance what is the most important to gather [78]. The power of visual sociograms can benefit many PER studies, especially with a qualitative or mixed-methods focus where the shared discussion space of the network can draw out additional nuance and detail.

## IV.C. Network visualization

Network data present special opportunities and challenges when it comes to visualization. Molina et al. [81] review ways that network visualizations can contribute to data collection and analysis. Their discussion emphasizes how visuals can highlight patterns and raise questions for further investigation, but need to be accurate. Work by Hogan and collaborators [79] has re-popularized sociograms in data collection, especially as web-based tools become more robust. Tubaro [75] discusses examples of visual methods at different data collection and analysis stages, and Rios-Aguilar and Deil-Amen [55] provide an interesting variant using anonymized roles of alters to show and discuss different patterns in students' social capital.

Molina et al. [81] find that most visualizations of personal networks are for exploratory purposes, but they can also be used for confirmatory analysis of conjectures (e.g., what network patterns would our hypotheses predict?), for model validation, or for post hoc analysis. Bidart et al. [82] developed a taxonomy of personal networks by visually sorting first, then testing a decision tree of structural measures to capture the observed types. Representations can range from reproducing participants' own placement choices [48,75], to computer-generated layouts that show areas of closeness and interconnection [75,82], to intentional abstraction and rearrangement for the purpose of highlighting group ties or power relations [83,84]. This is an evolving area in network analysis, and the mathematical focus on large-graph visualizations has a largely disjoint literature from the more sociological tradition of SNA. Physics education researchers might find tools of interest in both spaces, but here we have focused on resources especially relevant for personal network analysis, as those are less known in PER.

Egocentric network studies with more than a few participants have their own set of challenges, as displaying large numbers of sociograms may be visually overwhelming. One common approach is to select a few examples to show key trends [20,48]. Alternately, Brandes et al. [84] develop a method for visually summarizing networks in studies where the alters can be divided into distinct classes. By condensing information about alter classes' sizes and densities of connection, many graphs can be compared to look for trends [81,84,85], or aggregate graphs can be used to summarize and compare variability between ego groups of interest. This is very much a developing area of network studies where PER could add valuable insights.

# V. Examples from the literature

Because of the dominance of whole-network and quantitative approaches in PER, we have drawn most of our examples in this paper from outside the field. Additionally, to illustrate the theoretical frameworks or methods of interest, we used many sources in the broader sociology literature beyond education. We thus want to end by returning the focus to educational contexts.

Table 1 collects all references from this paper that take place in an educational setting (physics or otherwise) and use one or more of egocentric network analysis, qualitative methods, or mixed methods. Below we discuss three papers in more detail, particularly their problem of interest, their methods and use of theory, and how the network approach informed their results.

*Table 1: Education network studies referenced in this paper that use egocentric, qualitative, or mixed-methods designs. Since mixed-methods studies may have a completely quantitative network component, we have flagged qualitative network methods specifically.*

| Authors (year) | Title | Egocentric | Qualitative network methods | Mixed-methods design |
|---|---|---|---|---|
| A. E. N. Atherley, L. Nimmon, P. W. Teunissen, D. Dolmans, I. Hegazi, and W. Hu (2021) | Students' Social Networks Are Diverse, Dynamic and Deliberate When Transitioning to Clinical Training [48] | Y | Y | Y |
| R. M. Goertzen, E. Brewe, and L. Kramer (2013) | Expanded Markers of Success in Introductory University Physics [20] | Y | Y | N |
| C. Rios-Aguilar and R. Deil-Amen (2012) | Beyond Getting in and Fitting in: An Examination of Social Networks and Professionally Relevant Social Capital among Latina/o University Students [55] | Y | Y | N |
| L. Nimmon and A. Atherley (2022) | Qualitative ego networks in health professions education: Capturing the self in relation to others [7] | Y | Y | N |
| M. P. Skahill (2002) | The Role of Social Support Network in College Persistence among Freshman Students [49] | Y | N | N |
| G. Lorenz, Z. Boda, Z. Salikutluk, and M. Jansen (2020) | Social Influence or Selection? Peer Effects on the Development of Adolescents' Educational Expectations in Germany [60] | Y | N | N |
| O. Casquero, R. Ovelar, J. Romo, M. Benito, and M. Alberdi (2016) | Students' Personal Networks in Virtual and Personal Learning Environments: A Case Study in Higher Education Using Learning Analytics Approach [85] | Y | N | N |
| D. G. Wu, A. B. Heim, M. Sundstrom, C. Walsh, and N. G. Holmes (2022) | Instructor Interactions in Traditional and Nontraditional Labs [19] | Y | N | N |
| S. Dawson (2010) | 'Seeing' the Learning Community: An Exploration of the Development of a Resource for Monitoring Online Student Networking [41] | Y | N | N |
| M. de Laat, V. Lally, L. Lipponen, and R.-J. Simons (2007) | Investigating Patterns of Interaction in Networked Learning and Computer-Supported Collaborative Learning: A Role for Social Network Analysis [44] | N | Y | Y |
| J. Pulgar, C. Candia, and P. M. Leonardi (2020) | Social Networks and Academic Performance in Physics: Undergraduate Cooperation Enhances Ill-Structured Problem Elaboration and Inhibits Well-Structured Problem Solving [21] | N | N | Y |
| A. J. Gonsalves and H. R. Chestnutt (2020) | Networks of Support: Investigating a Counterspace That Provides Identity Resources for Minoritized Students in Post-Secondary Physics [30] | N | N | Y |
| A. Martínez, Y. Dimitriadis, B. Rubia, E. Gómez, and P. de la Fuente (2003) | Combining Qualitative Evaluation and Social Network Analysis for the Study of Classroom Social Interactions [42] | N | N | Y |

## V.A. Qualitative egocentric study of Latino/a social capital

Qualitative egocentric studies are well suited to exploring the nuances of personal situations and the mechanisms that may underlie larger network patterns. We chose this paper [55] as an example because it is a fully qualitative study that uses the visual power of sociograms to illustrate and compare profiles uncovered by the analysis.

The paper uses a qualitative egocentric study of 261 first-year Latina/o college students to understand the networks involved in their entry to college. The authors uncovered four broad profiles with implications for future support and DEI efforts. The study pursued two research questions: (1) What are the patterns in Latina/o students' social ties during their transition into and through the 1st year in college, and (2) What is the size of the network and the content of the exchanges between the Latina/o students and each network tie? Initial data were collected through 261 essays written by Latina/o students in a summer Bridge program. Subsequently, 61 of the students participated in semi-structured interviews, and the data collected through both methods were used to compose sociograms of students' networks.

The analysis proceeded from open coding, to generating ego networks using alter categories, to a second round of coding that incorporated network reflections. From this, four key network patterns emerged for discussion. The results highlighted the guidance Latino/a students receive regarding what and where to go to for college and the networks they mobilized trying to find their place in college (what the authors describe as "fitting in"). The paper goes beyond the idea of fitting in to also examine the networks students build to support their trajectories after college, from students to professionals. An interesting finding is that during college, Latina/o students do not build a network that can support this transition, and heavily rely on the networks they built when getting in and fitting in, even when their families don't have resources to provide career guidance.

The idea permeating the research is that the challenge is not only to enroll more students from diverse backgrounds and with minoritized identities. Understanding how those students get in and fit in is important, but also understanding the different ways students fit in an environment and how/if that connects to their careers after that stage. Also, understanding how students fit in can help DEI initiatives contribute to their career success. Getting in is important because it is a big move towards a professional career. Fitting in is important because the students are otherwise less likely to strive in that environment, but there is a whole career beyond that which is impacted by the networks those students built. In this sense, SNA represents an appropriate methodological choice.

When it comes to how the findings of this paper are presented, there are several ideas that can be introduced in PER by researchers interested in egocentric SNA. One idea is using roles instead of names to represent the alters, and including multiple egos in the same sociograms to visually contrast how egos build their networks. Abstracting individual alters into categories (e.g. family, clubs and organizations) in the first round of coding made it possible to find broad trends among the roles that were present or absent in students' networks. It also facilitated placing discovered network patterns on the same sociogram at once, providing some of the "big picture" overview that a whole network study might give.

Another interesting choice was to select the diagrams and ties that were relevant in different contexts, allowing comparison not only among egos but for the same egos in different moments or for different types of ties. The detail of the data allowed the researchers to contrast network patterns between the high school, college, and professionally relevant sets of ties. Both the qualitative and egocentric elements were essential to the analysis, where a whole-network approach did not match the study population, and a quantitative design would have pre-imposed researcher categories rather than starting from the students' experiences.

## V.B. Qualitative whole-network study of distance learning students

Although our paper focuses on egocentric network analysis, qualitative and mixed-methods designs can also bring new perspectives to whole networks. We chose an analysis by de Laat et al. [44] as a fairly uncommon case of qualitative whole-network analysis in a mixed-methods study, including a good discussion of how network analysis can be used with other methods to triangulate results.

Network analysis has been used in many studies of online or distance learning, often drawing on frameworks from computer supported collaborative learning (CSCL; [15]). The study by de Laat and collaborators focuses on a networked learning community of master's students in an E-learning program [44]. The researchers' goal was to discuss how to use SNA to study computer-supported collaborative learning and to add a layer of relational analysis to their existing case study data. The larger project used content analysis and interviews to describe teaching and learning in a distance learning course. Content analysis was used on a sample of the online messages from a 10-week workshop period to characterize learning and tutoring activities. Messages were coded under two classifications: one for different types of "on the task" work by students, and one focused on tutoring activities "around the task." This content analysis was used to extract patterns, and then critical event recall interviews were conducted with participants to gain feedback about those patterns and the context in which they arose. Then, qualitative analysis was used on the whole learning community network, using the time change in measures such as network density and centralization to think about the whole-class context for individual results from the earlier analyses.

The authors used analytic frameworks from CSCL to guide their content analysis. This use of theory structured the interaction patterns they looked for. These coded interactions were then interpreted using time-sequenced network diagrams to understand how the learning community evolved at the whole-class level. The earlier stages of the analysis identified patterns and participants' explanations of their learning processes. Network analysis added a view of how participation changed over time, with people becoming more central or peripheral at different stages of the 10-week course project. Combining network and other data also highlighted that students could be central for different reasons (e.g. debating and putting forward new ideas vs. managing group activity).

Qualitative analysis can be difficult to perform on larger networks, but many departments will have courses of a size that would fit well with this kind of design. With hybrid and online courses increasingly common, the diverse CSCL network literature [15,86] has many examples to offer physics education researchers, including some egocentric cases [85].

## V.C. Mixed-methods egocentric study of medical students

The final case we highlight is a mixed-methods egocentric study of support networks in medical students transitioning into clinical clerkships [48]. These transition periods are significant parts of medical education and professionalization, but are under-studied from a social perspective. We chose this paper as an example of a mixed-methods study where both quantitative and qualitative tools were used to interpret the networks.

The research questions were: (1) Who are in the social networks of undergraduate students transitioning to the clinical environment? (2) To what extent do undergraduate students' networks change as they transition to the clinical environment? What are the mechanisms behind these changes? Theoretically, the authors drew on social constructivism, social support, and communities of practice. The SNA data were collected by interviews, including participant sociogram construction, at the beginning and four-month point of the clerkships. Interviews solicited the name, role, gender, and polarity (positive or negative interactions) for everyone who impacted the participant's transition. The analysis compared network size and role composition between time points and found that the networks did not significantly grow, but the distribution of roles changed, with ties shifting away from nurses and near-peers (students in other years) and toward doctors and peers (students in the same cohort).

These patterns were explored through inductive thematic analysis of the transcripts, sociograms, and videos recorded from interviews. The qualitative analysis shed light on the mechanisms behind the changes in networks, such as the persistence of ties to doctors who integrated students into the clinical team, or the time it took to build trust in peers and start sharing professional norms. For physics education researchers, this work gives a mixed-methods example where the network framing was integrated at all stages of data collection, analysis, and interpretation of results.

# VI. Conclusions

Most network analyses in PER have been quantitative studies focused on network function and structure or student persistence in physics. In this paper, we argue that PER would benefit from—and can contribute to—the wider range of tools, frameworks, and research questions found in educational network research. In particular, egocentric approaches offer many options for extending beyond the boundaries of a single course or for considering widely separated individuals through network perspectives. Qualitative or mixed-methods approaches allow additional depth and nuance in understanding the mechanisms underlying network ties, which are generally invisible in quantitative studies. In this review, we have tried to highlight both challenges and areas of potential, as well as provide many resources and examples to help researchers explore further.

We organized the paper around key questions that arguably could benefit any network study, but which we found especially relevant in mapping out a methodologically unfamiliar space. The network literature spreads over many disciplinary journals, so we prioritized key results and reviews over closeness to physics. Finally, we end with a few meta-recommendations to those who might want to expand into these types of analyses:

- Answer the "big questions" posed above not just in study design, but when reporting results. Because there are so many possible orientations toward networks, being explicit helps other researchers to understand claims and to highlight gaps for possible future study.
- Be clear and descriptive about methodological choices. In mixed-methods network studies, the purpose of the "mixing" is not always clear [12], and post-data-collection processing is often under-described. Studies using less-common tools like essays [55] to uncover network information can contribute methodologically by documenting how they go from the original form of the data to encoded network ties.
- Decide early how much standardized or quantified comparison between networks is desired. For detailed quantitative study, the sequence and wording of prompts should be as uniform as possible [77]. However, the freedom allowed by less-structured interview methods might be a higher priority depending on the research questions.

Much work remains to be done with social network analysis in PER, and we look forward to seeing how these tools can be applied to the problems of teaching, learning, and practicing physics.

# Acknowledgments

Justin Gutzwa and Lily Thomas of the PERU research group have contributed significantly to the larger project this paper is part of. This work was supported in part by National Science Foundation awards DUE-2100024, 2055237, and 2054920.